\documentclass[1p]{elsarticle}

\usepackage{lineno,hyperref}

\journal{Icarus}

\biboptions{authoryear}

\begin{document}

\begin{frontmatter}

\title{Gas flow in Martian spider formation}

\author[stir]{Nicholas Attree\fnref{myfootnote}}
\ead{n.o.attree@stir.ac.uk}
\author[stir,iwf]{Erika Kaufmann}
\author[ltu,stir]{Axel Hagermann}
\address[stir]{Faculty of Natural Sciences, University of Stirling, UK}
\fntext[myfootnote]{Corresponding author}
\address[iwf]{Now at Institute for Space Research Graz, Austrian Academy of Sciences, Austria}
\address[ltu]{Lule{\aa}  University of Technology, Space Campus Kiruna, Sweden}

\begin{abstract}
Martian araneiform terrain, located in the Southern polar regions, consists of features with central pits and radial troughs which are thought to be associated with the solid state greenhouse effect under a CO$_{2}$ ice sheet. Sublimation at the base of this ice leads to gas buildup, fracturing of the ice and the flow of gas and entrained regolith out of vents and onto the surface. There are two possible pathways for the gas: through the gap between the ice slab and the underlying regolith, as proposed by \citet{Kieffer2007}, or through  the pores of a permeable regolith layer, which would imply that regolith properties can control the spacing between adjacent spiders, as suggested by \citet{Hao}. We test this hypothesis quantitatively in order to place constraints on the regolith properties. Based on previously estimated flow rates and thermophysical arguments, we suggest that there is insufficient depth of porous regolith to support the full gas flow through the regolith. By contrast, free gas flow through a regolith--ice gap is capable of supplying the likely flow rates for gap sizes on the order of a centimetre. This size of gap can be opened in the centre of a spider feature by gas pressure bending the overlying ice slab upwards, or by levitating it entirely as suggested in the original \citet{Kieffer2007} model. Our calculations therefore support at least some of the gas flowing through a gap opened between the regolith and ice. Regolith properties most likely still play a role in the evolution of spider morphology, by regolith cohesion controlling the erosion of the central pit and troughs, for example.
\end{abstract}

\begin{keyword}
Mars; Mars, surface; Mars, polar geology: ices
\end{keyword}

\end{frontmatter}

\section{Introduction}

Araneiform terrain consists of groups of so--called spiders; radial, dendritic arrangements of troughs connecting to central pits, found at southern high latitudes of Mars \citep{Kieffer2000}. Their occurence was thought to be limited to the South Polar Layered Deposits (SPLD). Spiders were also discovered outside the area of the SPLD, although the extent of the SPLD was subsequently re--defined. \citet{Tanaka2014} or \citet{Schwamb2017} provide more insight the geography of spiders and the SPLD. These features are perennial, visible all year round, scoured up to several metres deep into the regolith, and can be tens to hundreds of metres in size \citep{Hansen2010}. A slab of CO$_{2}$ ice begins to form in the autumn and covers this terrain during the southern winter and spring \citep{Kieffer2000}. Based on this hypothesis, \citet{Piqueux2003} were the first to define and map the distribution of the spider features, confirming the correlation of spiders with the existence of highly transparent CO$_{2}$ slab ice, which overlies the largely unconsolidated particulate regolith of the South Polar Layered Deposits.

Over the years, what is now commonly referred to as the Kieffer model has undergone extensive extensions and refinements, but there is general consensus that spiders are formed by gas flowing under the CO$_{2}$ ice  \citep{Piqueux2003, Kieffer2006, Kieffer2007, Piqueux2008, Portyankina2010, Thomas2010, Pilorget2011, Thomas2011, Martinez2012, Pilorget2016}. Sub--ice gas production is enabled by the transparency of $CO_2$ ice (see e.g. \citet{Hansen2005} and references therein). The ice slab allows spring--time insolation to penetrate to its base and warm it in a solid-state greenhouse, a phenomenon similar to the atmospheric greenhouse effect \citep{Brown1987, Fanale1990}. This means that sunlight can penetrate the CO$_{2}$ ice layer down to a dust deposit at depth within or below the ice, where the radiation is absorbed. The temperature increase here leads to sublimation of the CO$_{2}$ ice on the boundary between the ice layer and the underlying material. Sublimating gas then builds up until the pressure fractures the ice, opening a vent to the surface, followed by gas flow from the surrounding area into the vent and ejection. Regolith particles become entrained in the flow, being ejected along with the gas,  whilst also eroding the troughs and pits \citep{Thomas2010}. Larger and heavier grains end up forming dark fans on top of the ice. These can be more radial or unidirectional, depending on local surface winds carrying the material over the top of the ice sheet. whilst the smallest and lightest grains are ejected into the atmosphere \citep{Portyankina2010}. This process is thought to repeat seasonally, so spiders play an important role in Mars' global dust cycle and therefore the planet's climate (see also \citep{Kieffer2000, Piqueux2003}). Over time, spiders can effect substantial morphological change and they are representative of a type $CO_2$--sublimation associated erosion representative for Mars. In fact, most morphological changes on Mars are considered to be $CO_2$ related today \citep{Piqueux2008,Hansen2005}.

As for the gas trapped unerneath the ice slab, there are several escape routes to consider. In the commonly accepted Kieffer model  \citeyear{Kieffer2007},  gas pressure physically lifts the ice slab, levitating at least part of the overlying slab and allowing gas to flow through an ice--regolith gap. This allows a radially converging flow to a central vent to drain a large area of sublimating ice through only a small ($\sim$cm thick) gap.
However, we need to bear in mind the porous nature of Mars' regolith, which might permit gas to flow within the substrate underlying the ice. The importance of the latter gas escape route has recently been stressed by \citet{Hao}, who established a link between spider spatial density and regolith morphology. They observed that spiders are grouped at mutual distances smaller than that of random spacing, suggesting a controlling length--scale whereby a spider inhibits the formation of new spiders out to some distance from itself. Since this distance was seen to vary between different classes of spider morphology, \citet{Hao} proposed that it is controlled by the regolith properties, e.g.~that different regolith porosities/permeabilities constrain the gas flow by different amounts, and therefore the radius around each spider where gas can build up. Evidence for the relevance of regolith morphology was observed by \citet{Hao2020}. These recent results highlight that the potential effectiveness of gas flow through the regolith itself --- rather than through a regolith--ice gap --- has to be addressed because it seems vital for our understanding of spiders.

We approach the problem by providing suitable limits on the effectiveness of the underground flow of gas, based on constraints on regolith properties such as porosity/permeability combined with some known gas physics. Regolith porosity is a function of depth and therefore our treatment of the problem has to involve some stratigraphic considerations. We derive the depth of porous regolith needed to support the magnitudes of flow rate previously calculated in hydrodynamics simulations \citep{Thomas2011b, Thomas2011} and energy balance models \citep{Kieffer2007, Pilorget2011, Pilorget2016}. We then compare this to the required gap size to support laminar free--flow between the ice and regolith, and to the expected bending of the ice sheet in response to the rise in gas pressure beneath. We present the calculations in Section \ref{modelling}, before discussing the results and their implications in Section \ref{discussion}, and concluding in Section \ref{conclusion}.

\section{Modelling}
\label{modelling}

\begin{figure}
\noindent\includegraphics[width=\textwidth]{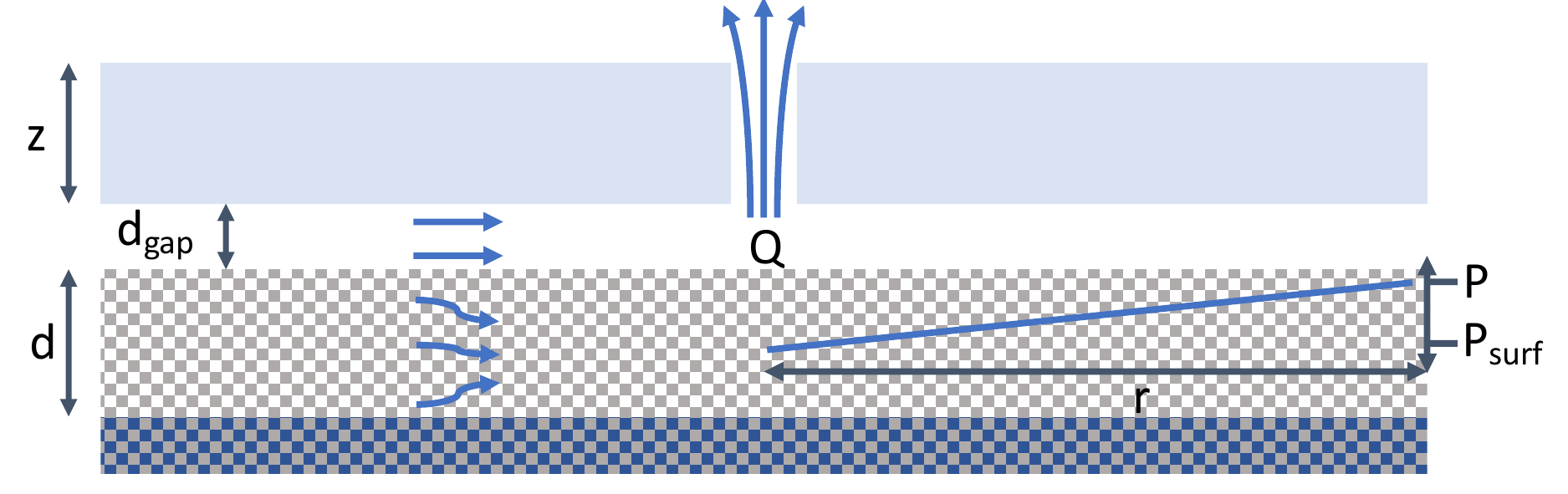}
\caption{Conceptual model for spider formation based on a layer of transparent CO$_{2}$ ice of thickness $z$},  overlaying permeable regolith of thickness $d$, above an impermeable ice--rich regolith layer shaded dark. Gas pressure in the regolith increases from near--atmospheric pressure around the spider vent to a maximum value at the edge of the spider. Some models, such as those by  \citet{Hao} and \citet{Pilorget2011} assume no gap between ice and regolith, $d_{gap} \to 0$. See section \ref{gaspressure} for a detailed description of parameters used.
\label{fig_model}
\end{figure}

The most common model for spider formation is illustrated in Figure \ref{fig_model}; CO$_{2}$ slab ice condenses on-top of regolith over the Martian winter. When insolation begins to penetrate this transparent ice layer in spring, CO$_{2}$ ice at the base of the slab and top of the regolith starts to sublimate, raising the gas pressure to the point where it exceeds the strength of the ice slab and fracturing occurs. In the model of \citet{Kieffer2007}, this gas pressure levitates the slab above the regolith surface by a distance of $d_{gap}$, estimated at around one or two centimetres. Gas can also diffuse into the permeable part of the regolith \citep{Hao}, which extends down to some depth $d$ (also on the order of centimetres \citep{Kieffer2007, Portyankina2010}), below which the regolith is made impermeable by water ice filling the pore spaces.  See \cite{Mellon2004} for a detailed discussion of the geography of ground ice on Mars. Irrespective of whether the gas flows above or through the regolith to begin with, once fracturing has opened a vent to the surface, a pressure gradient across the radius of the spider drives gas flow to it, either as free--flow through the gap or viscous flow through the porous regolith. In the latter case, flow is limited by the pressure gradient and the regolith permeability, and we estimate these below.

\subsection{Gas pressure}
\label{gaspressure}

The sublimation rate at the bottom of the ice slab controls both the pressure build-up and the flow rate into the spider vent. Upon initial fracturing there may be a very energetic flow, driven by the large initial pressure gradient, followed by a reduction to a steady state balance between gas production and ejection. Both these cases were simulated with a commercial hydrodynamics code in \citet{Thomas2011b} and \citet{Thomas2011}.

The gas production rate is estimated by comparing the amount of energy deposited by insolation at the ice-regolith boundary with the latent heat of sublimation of CO$_{2}$. \citet{Kieffer2007} assumed a perfectly transparent ice slab and calculate an area gas production rate of $\dot{m}=1.157\times10^{-4}$ kg m$^{-2}$ s$^{-1}$. \citet{Thomas2011} meanwhile assume $25\%$ of the solar energy penetrates a $z=20$ cm thick slab (based on \citet{Pilorget2011}) and arrive at $\dot{m}=2.21\times10^{-5}$ kg m$^{-2}$ s$^{-1}$.

This constant gas production during spring--time daylight hours cannot escape and swiftly fills up all available sub-ice volume. Gas pressure will at this point be equal to the cryostatic pressure of $P=P_{surf} + \rho_{ice}~gz$, where $g=3.72$ m s$^{-2}$ is the Martian gravity and $\rho_{ice} = 1606$ kg m$^{-3}$ is the solid CO$_{2}$ ice density. Taking $z$ as 0.7 m, as in \citet{Pilorget2011}, and $P_{surf}\approx600$ Pa, results in sub-ice pressures of $P_{cryo}\approx4700$ Pa. This pressure will continue building until the stress in the ice slab above exceeds its rupture strength and it fractures. During the initial rupture, \citet{Thomas2011b} simulated a total flow rate of $Q_{i}=0.05-0.5$ kg s$^{-1}$ draining a 400 square metre area into a central vent. Subsequently, ejection of gas reduces the pressure gradient and, in a steady state, the 400 square metre area will produce a total flow some $\sim5$ times smaller: $Q_{s}=0.00884$ kg s$^{-1}$ for $\dot{m}=2.21\times10^{-5}$ kg m$^{-2}$ s$^{-1}$. Hydrodynamics simulations of these magnitudes of flow rate produce gas and dust ejection velocities of $\sim10$ ms$^{-1}$, consistent with the erosion of regolith and the spread of fine material in fans on the surface \citep{Thomas2011b, Thomas2011}.

A gas production area of 400 square metres corresponds to a circular gas production radius of only $r_{P}=11.3$ m, somewhat smaller than the  55 m separation (spider radius $R\sim25$ m) between densely packed spiders found in \citet{Hao}. Gas flows here therefore represent lower limits; larger spiders will have larger mass flow rates, but the ice breaking pressure will remain the same. 

\begin{figure}
\noindent\includegraphics[width=\textwidth]{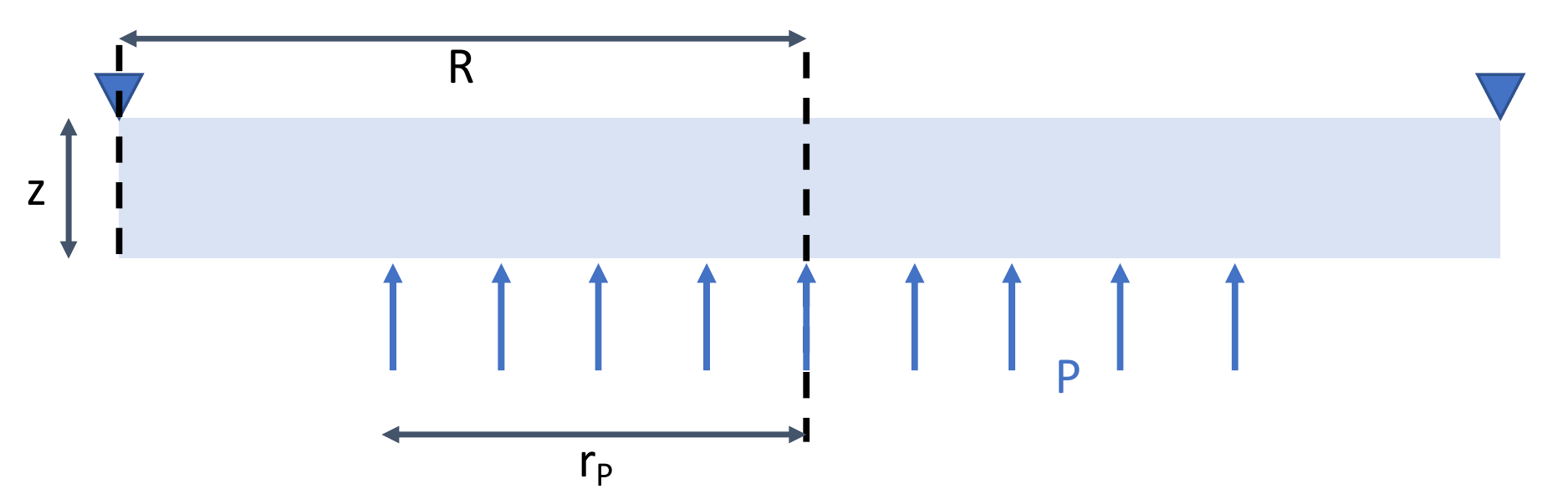}
\caption{Model for the breaking of a thin plate as in \citet{Portyankina2010}. Pressure $P$ builds up in a circular region of radius $r_{P}$ under a plate of thickness $z$ and radius $R$ with simply supported edges.}
\label{fig_model2}
\end{figure}

\begin{figure}
\noindent\includegraphics[width=\textwidth]{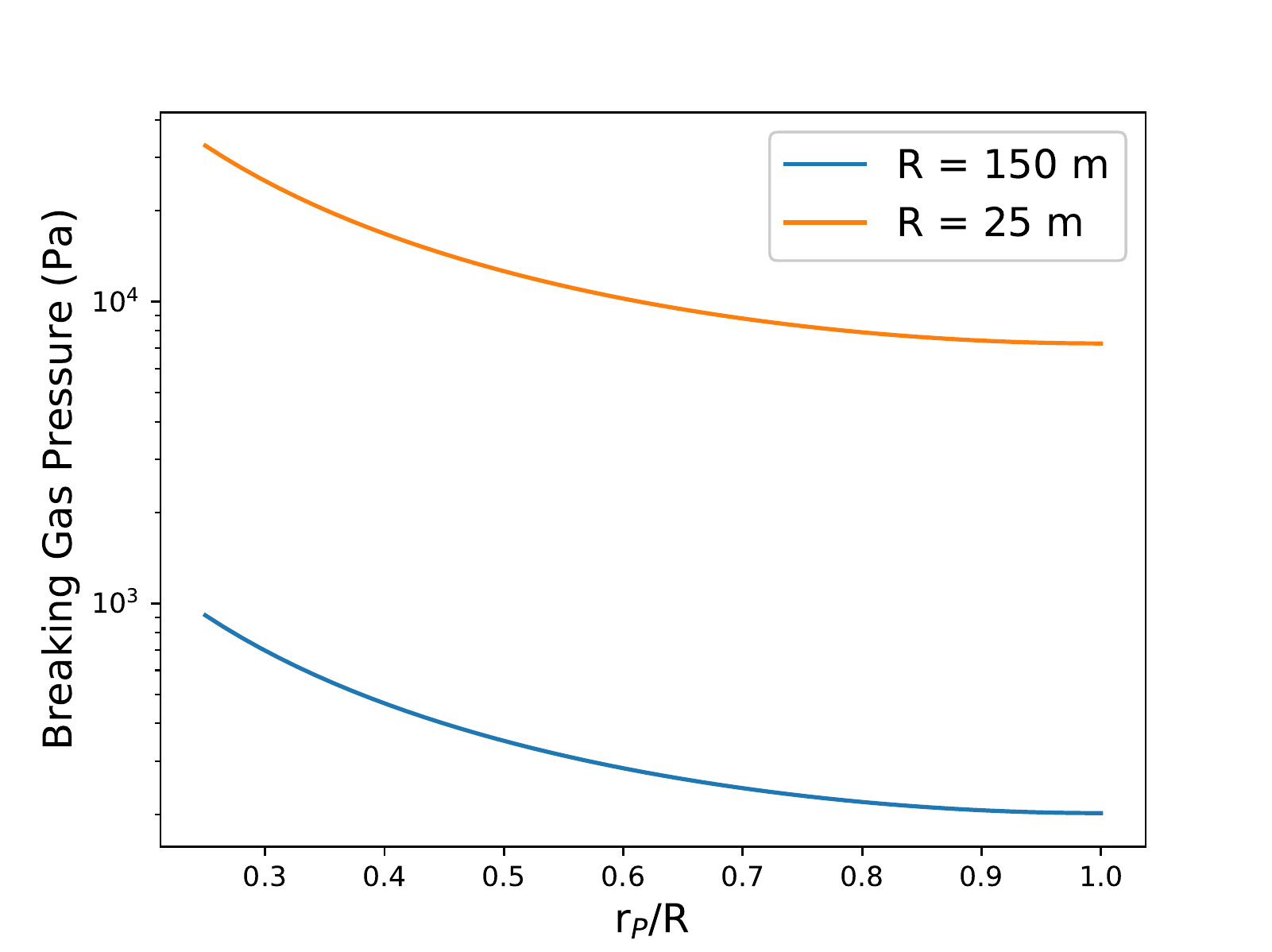}
\caption{Gas pressure required to fracture a sheet of CO$_{2}$ ice of thickness $z=0.7$ m and radius $R$ for different scales of gas buildup area $r_{P}$.}
\label{fig_pcrit}
\end{figure}

The critical breaking pressure can be estimated using the fracturing model of \citet{Portyankina2010}, which is based on an engineering model of the bending of a thin sheet, as shown in Figure \ref{fig_model2}. \citet{Portyankina2010} give the maximum stress at the middle of a sheet of radius $R$, and thickness $z$, when subjected to a constant pressure $P$ over a radius of $r_{P}$. Recently, \citet{Kaufmann2019} have measured the yield strength of CO$_{2}$ ice at the relevant temperature as $\sigma=12.3$ MPa, and so we can set the stress equal to this value and solve for the critical gas pressure needed to cause fracturing
\begin{linenomath*}
\begin{equation}
P_{crit} = \frac{8\sigma z^{2}}{3r_{P}^{2}}\left[4 - (1 - \nu) \frac{r_{P}^{2}}{R^{2}} + 4(1 + \nu) \ln\frac{R}{r_{P}}\right]^{-1}.
\label{pcrit}
\end{equation}
\end{linenomath*}
Here, $\nu_{I}=0.544$ is the Poisson's ratio of CO$_{2}$ ice \citep{Yamashita}. It is assumed that the pressure builds up and is constant over a length--scale $r_{P}$, comparable to the spider radius $R$ (see Fig.~\ref{fig_model2}). We favour a ratio of $r_{P}/R$ equal to one, in contrast to the value of 0.25 as used by \citet{Portyankina2010}. We think this is more realistic because gas pressure should build up over the whole radius of the spider, right up to its edge where the maximum sub-ice pressure exists, and beyond-which pressure begins to decrease again with decreasing distance towards the next spider (Fig.~\ref{fig_model}). Nevertheless, we plot $P_{crit}$ for a range of ratios and for two different overall spider sizes in Figure \ref{fig_pcrit}.

It can be seen from  Figure \ref{fig_pcrit} that large pressures, exceeding the cryostatic pressure of $P_{cryo}\sim5$ kPa, can develop before fracturing occurs. This is especially the case for the smaller spiders (small R) and for small gas production radii (small $r_{P}/R$), both of which can be understood as reducing the total area that the pressure is applied to, meaning more gas pressure must be added to achieve the same total stress in the ice. For the densely packed spider separations observed in \citet{Hao} and \citet{Hao2020} and a gas production radius of $r_{P}=0.25 R\sim6.5$ m (even smaller in size than the case simulated by \citet{Thomas2011} above), a total pressure of $P_{crit}\approx33$ kPa can build up before fracturing occurs. Thus, in this case our maximum pressure gradient driving gas flow at the initial breach would be $(P_{crit}-P_{surf})/r_{P}\approx33$ kPa$/6.5$ m.

\subsection{Permeability}

The permeability of a porous medium can be expressed as (see for example \citet{Balme})
\begin{linenomath*}
\begin{equation}
K = \frac{D^{2}\phi^{3}}{72\tau(1-\phi)^{2}},
\end{equation}
\end{linenomath*}
where $D$ is the constituent particle size, $\phi$ is porosity and $\tau$ is tortuosity. Typical values used for Martian regolith are $\phi=40\%$, $\tau=25/12$ \citep{Balme,Demidov2015}, and particle sizes between $\sim$microns and a millimetre (e.g. \citeauthor {Morgan2018}, \citeyear{Morgan2018}). Over this range of particle sizes $K$ changes by several orders of magnitude, exceeding any uncertainty in $\phi$ and $\tau$. For gasses at a low pressure $P$ such as we are dealing with (rather than liquid flows), K must be corrected by the `Klinkenberg correction' \citep{Klinkenberg}
\begin{linenomath*}
\begin{equation}
k_{G} = k + \frac{6.9}{P}k^{0.64},
\end{equation}
\end{linenomath*}
where pressure is expressed in psi and permeability in mdarcies. Conversion back to SI units is performed using $k = cK$ and $k_{G} = cK_{G}$, with $c=0.986923\times10^{-3} \mu$m$^{2}$.

With the above permeability, Darcy's law for the total mass flow, $Q$, through a cross-sectional area $A$ of porous medium, driven by a pressure gradient of $dP/dr$, is \citep{Ahmed2001}
\begin{linenomath*}
\begin{equation}
Q = -\frac{\rho K_{G}A}{\eta}\frac{dP}{dr},
\label{Darcy}
\end{equation}
\end{linenomath*}
with the gas density $\rho$ and viscosity $\eta$ given by
\begin{linenomath*}
\begin{equation}
\rho = \frac{Pm_{CO2}}{k_{B}T},
\end{equation}
\end{linenomath*}
and
\begin{linenomath*}
\begin{equation}
\eta = \rho \lambda \sqrt{\frac{2k_{B}T}{\pi m_{CO2}}},
\end{equation}
\end{linenomath*}
respectively. Here $k_{B}$ is Boltzmann's constant, $m_{CO2}=7.3\times10^{-26}$ kg and $d_{CO2}=330\times10^{-12}$ m are the molecular mass and diameter of CO$_{2}$, and the mean free path is given by
\begin{linenomath*}
\begin{equation}
\lambda = \frac{k_{B}T}{\sqrt{2} \pi P d_{CO2}}.
\end{equation}
\end{linenomath*}
It is assumed that gas temperature, $T$, is equal to the sublimation temperature at the sub-ice pressure
\begin{linenomath*}
\begin{equation}
T = \frac{-b}{\ln(P/100) - a},
\end{equation}
\end{linenomath*}
where we use the constants $a=23.3494$ and $b=3182.48$, as in \citet{Thomas2011b}. Under the 0.7 m deep slab these values are $T=163$ K (around 10 K higher than at the surface) $\rho=0.155$ kg m$^{-3}$, $\lambda=9.75\times10^{-7}$ m, and $\eta=2.12\times10^{-5}$ Pa s.

 Assuming a cylindrical geometry with a thickness of permeable regolith $d$ metres, the cross-sectional area through which gas can flow, $A$, at a distance $r$ from the central vent, is $A=2\pi rd$. The pressure gradient, $\frac{dP}{dr}$, can be approximated as $\Delta P/r_{P}$, where $\Delta P$ is the pressure drop from the high-pressure, sub-ice environment to the low-pressure vent over a length-scale $r_{P}$. Close to the central vent (i.e.~$r \ll R$), almost the entire gas flow, $Q$ from above, must pass through an area of permeable regolith. We therefore set $r=1$ m and obtain an expression from Eqn.~\ref{Darcy} for the required depth of regolith in terms of this flow as
\begin{linenomath*}
\begin{equation}
d = \frac{Q\eta}{2\pi \rho K_{G}}\frac{r_{P}}{\Delta P}.
\label{d}
\end{equation}
\end{linenomath*}
Note that we have assumed a constant permeability for the regolith layer while, in reality, porosity, and therefore permeability, is generally reckoned to decrease with depth (see for example \citet{Morgan2018}). Decreasing permeability will restrict gas flow rates meaning that, again, our estimate here is a `strongest-case' for flow through porous regolith.

The length--scale $r_{P}$ is the gas buildup radius of above and should be similar in size to the spider radius $R$. Indeed, \citet{Hao} suggest that the distance between adjacent spiders is determined by the pressure fall-off, so that new fractures open (forming a new spider) at a distance from the existing vent where pressure increases to it's maximum sub-ice value, $r_{P}\approx R$. In the steady state case, therefore, a further simplification can be made. Total flow into the vent is the sum of the per-metre gas production rate over the area of the spider $Q=\dot{m}\pi R^{2}$, and, having made the assumption $r_{P}=R$, Eqn.~\ref{pcrit} can be simplified and combined with Eqn.~\ref{d} to make the dependence on spider size $R$ explicit. We also make the additional assumption that $\Delta P \approx P_{crit}$ (neglecting the comparatively small surface pressure) so that we are using the maximum pressure gradient possible: the ice breaking pressure, rather than the cryostatic pressure. Then $R$ can be expressed as
\begin{linenomath*}
\begin{equation}
R = \left[\frac{16d\rho K_{G} \sigma z^{2}}{3\eta\dot{m}(3 + \nu)}\right]^{1/5}.
\label{L}
\end{equation}
\end{linenomath*}
This is the maximum radius of a cylindrical volume of porous regolith which can be effectively drained into a central vent at a steady state production $\dot{m}$ before gas pressure at the edge exceeds ice strength and a new vent opens.

\section{Results and discussion}
\label{discussion}

We compute the required regolith depth $d$ from Equation \ref{d} for a range of regolith particle sizes and two scenarios: steady state flow and just after the initial rupture. In the steady state, we use the minimum flow rate from above, $Q_{s}=0.00884$ kg s$^{-1}$, and cryostatic pressure over a length-scale equal to the smaller spider separations, i.e.~$\Delta P/r_{P}=(P_{cryo}-P_{surf})/r_{P}\approx5$kPa$/25$ m. For the rupture case we use the strongest case of $(P_{crit}-P_{surf})/r_{P}\approx33$ kPa$/6.5$ m, from above, as well as the slightly higher \citet{Thomas2011b} flow rate ($Q_{i}=0.05$ kg s$^{-1}$).

\begin{figure}
\noindent\includegraphics[width=\textwidth]{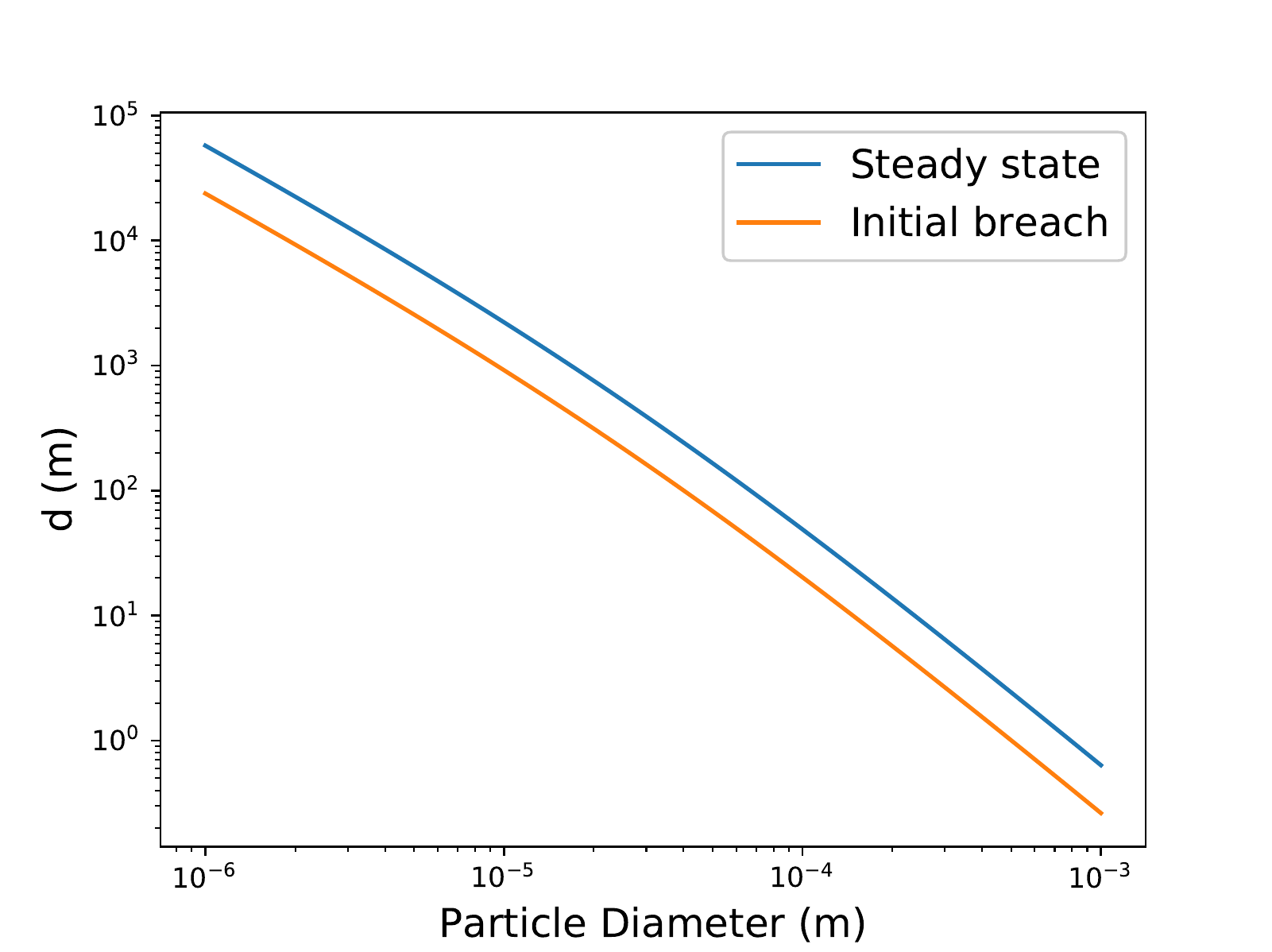}
\caption{Required depth of the permeable layer, $d$, to sustain the minimum mass flow, $Q$, for various particle diameters and pressure gradients.}
\label{results}
\end{figure}

Figure \ref{results} shows the results: in the steady state, a permeable regolith layer of thickness at least $d\approx13.4$ m is needed to support the flow rate for the likely particle size of up to $D\approx200~\mu$m \citep{Kieffer2007}. This is rather large, given the presence of ground ice and a decreasing regolith porosity with depth. Even in the case of the strongest possible pressure gradient just after rupture, the required thickness is still $d\approx5.4$ m. Only in the case of very large regolith particles ($\sim1$ mm) and the strongest pressure gradients does the required layer thickness fall below one metre. An impermeable layer of water-ice bonded regolith is generally assumed to be present in the Martian polar regions at a depth of between several centimetres \citep{Portyankina2010, Pilorget2011} and the seasonal skin depth of $\sim0.8$ m \citep{Kieffer2007}. This is supported by thermal inertia \citep{BanfieldFeldman} and gamma-ray and neutron spectrometry measurements \citep{Boynton81, diez2008}, as well as the recent detection of shallow-buried water ice cliffs at mid-latitudes \citep{Dundas199}.

\begin{figure}
\noindent\includegraphics[width=\textwidth]{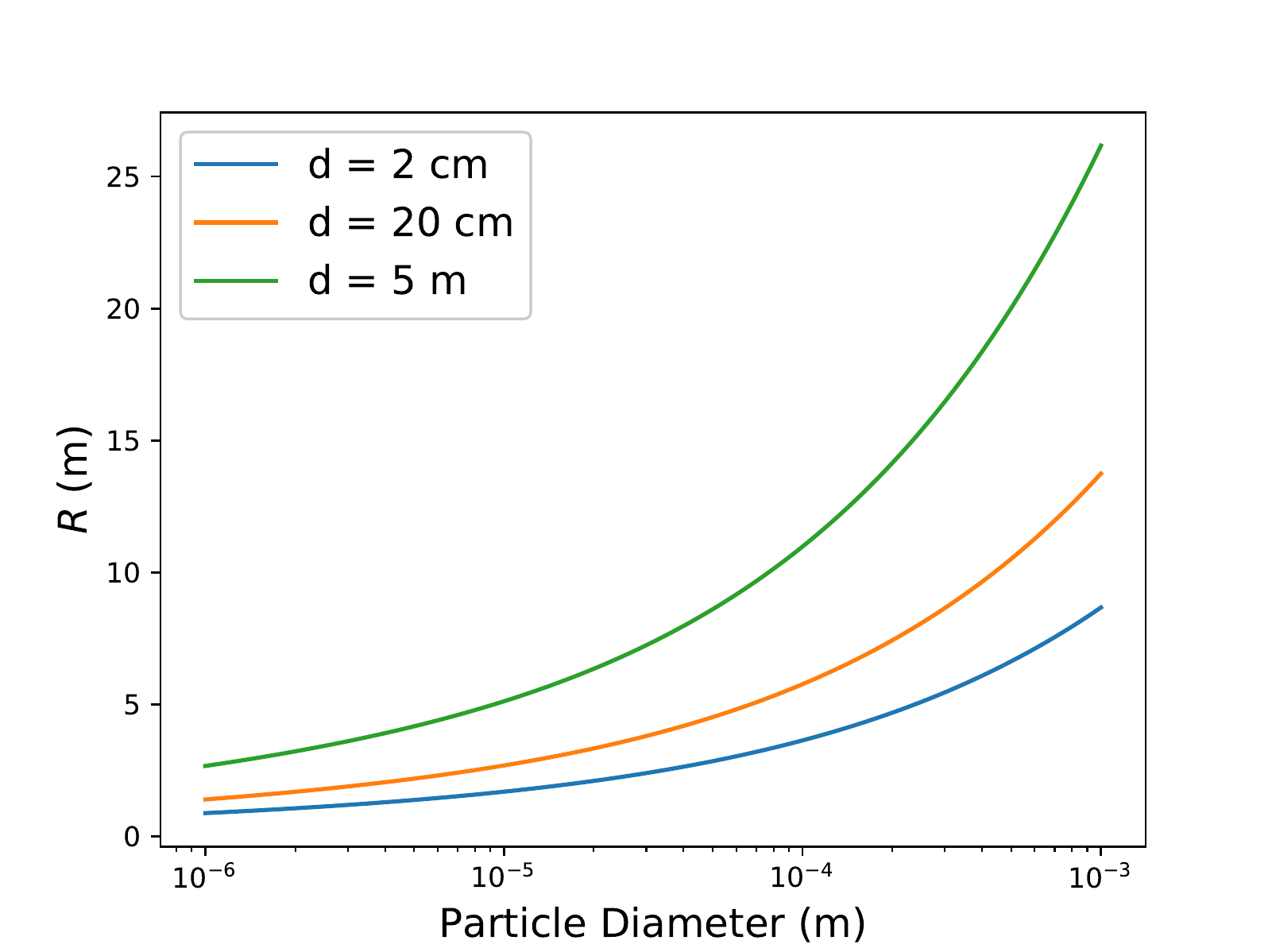}
\caption{Maximum radius $R$, over which steady-state gas production can be drained before pressure build--up exceeds the overlying ice strength at the edge, for different regolith depths.}
\label{results_L}
\end{figure}

Given the above, it seems unlikely that the total gas flow into a spider vent can occur purely through a porous regolith. What then is the largest size of a feature than can be supported this way? Figure \ref{results_L} shows the results of Equation \ref{L}, where we assumed a steady state flow with the minimum gas production and a fixed geometry, and solve for $R$ for a set of permeable regolith depths. This is the length-scale over which a spider can effectively drain gas to a central vent before pressure exceeds the overlying ice strength at the edge. Even in the case of large regolith depths, the length-scales ($\sim15$ m) are below the observed minimum spider separation distances ($\sim25$ m) for a particle size below $200~\mu$m. For more realistic porous layer thicknesses, only areas of a few metres in radius can be drained. Gas flow through porous regolith therefore seems incompatible with the expected polar stratigraphy. Even in the most amenable case of the smallest distances between araneiforms, with the highest pressure gradients and the lowest gas flow rates, there is insufficient depth of permeable regolith to support the flow associated with the whole spider area venting through a single central vent. Small areas of a few metres in radius around local vents (such as the so-called `baby spiders' identified by \citealt{Schwamb2017}) may be drained by viscous flow through the pores, but it seems likely that the larger (tens to hundreds of metres across) spiders must involve the original \citet{Kieffer2007} model of ice slab levitation and free gas flow.

\begin{figure}
\noindent\includegraphics[width=\textwidth]{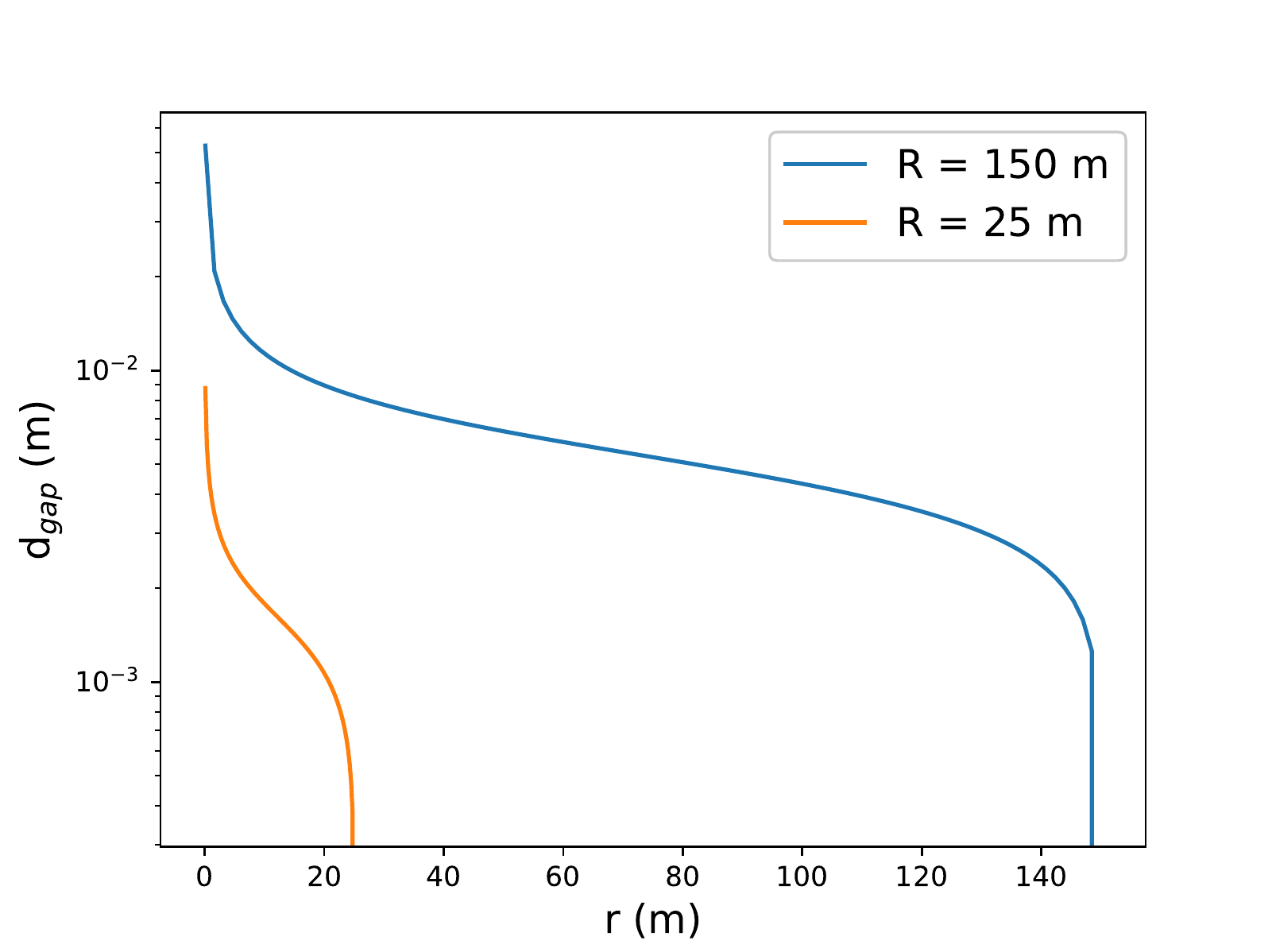}
\caption{Required gap size, $d_{gap}$, to support free-flow of the total gas production, between the ice and regolith, across the radii of two different spider sizes.}
\label{results_freeflow}
\end{figure}

Slab levitation is likely when considering the large gas pressures, in excess of cryostatic pressure, calculated above, as well as the small gap size needed for relatively large flows. This required gap size $d_{gap}$ can be computed. Assuming laminar Poiseuille flow between two parallel plates (an approximation that likely holds away from the energetic flow at the vent itself) and a pressure gradient of $\Delta P/R$, the mass flow rate per unit circumference is given by $q = \frac{\rho d_{gap}^{3}}{12\eta}\frac{\Delta P}{R}$. The total mass flow into an annulus at radius $r$ can also be expressed in terms of the steady state area production rate as $Q(r) = \dot{m}\pi(R^{2} - r^{2})$, so that per unit of circumference it is $\frac{\dot{m}}{2}(R^{2}/r - r)$. The two can then be equated and solved for $d_{gap}$, the required distance between the two plates to accommodate the flow, as
\begin{linenomath*}
\begin{equation}
d_{gap} = \left[\frac{6\dot{m}\eta}{\rho\Delta P}\left(\frac{R^{3}}{r} - rR\right)\right]^{1/3}.
\label{d_freeflow}
\end{equation}
\end{linenomath*}
Figure \ref{results_freeflow} shows the results of equation \ref{d_freeflow} for the two spider sizes from above. The required depth is greatest for the large mass flows of the bigger spiders, and increases towards the central vent as the gas is concentrated together in the converging flow (it is undefined at $r=0$, the vent itself). Nonetheless, the required gap between the regolith and ice, $d_{gap}$ is much smaller than the depth of permeable regolith $d$ calculated above and rarely exceeds one centimetre. Thus the whole flow can be supported by a gap of a few centimetres, as suggested by \citet{Kieffer2007}.

\begin{figure}
\noindent\includegraphics[width=\textwidth]{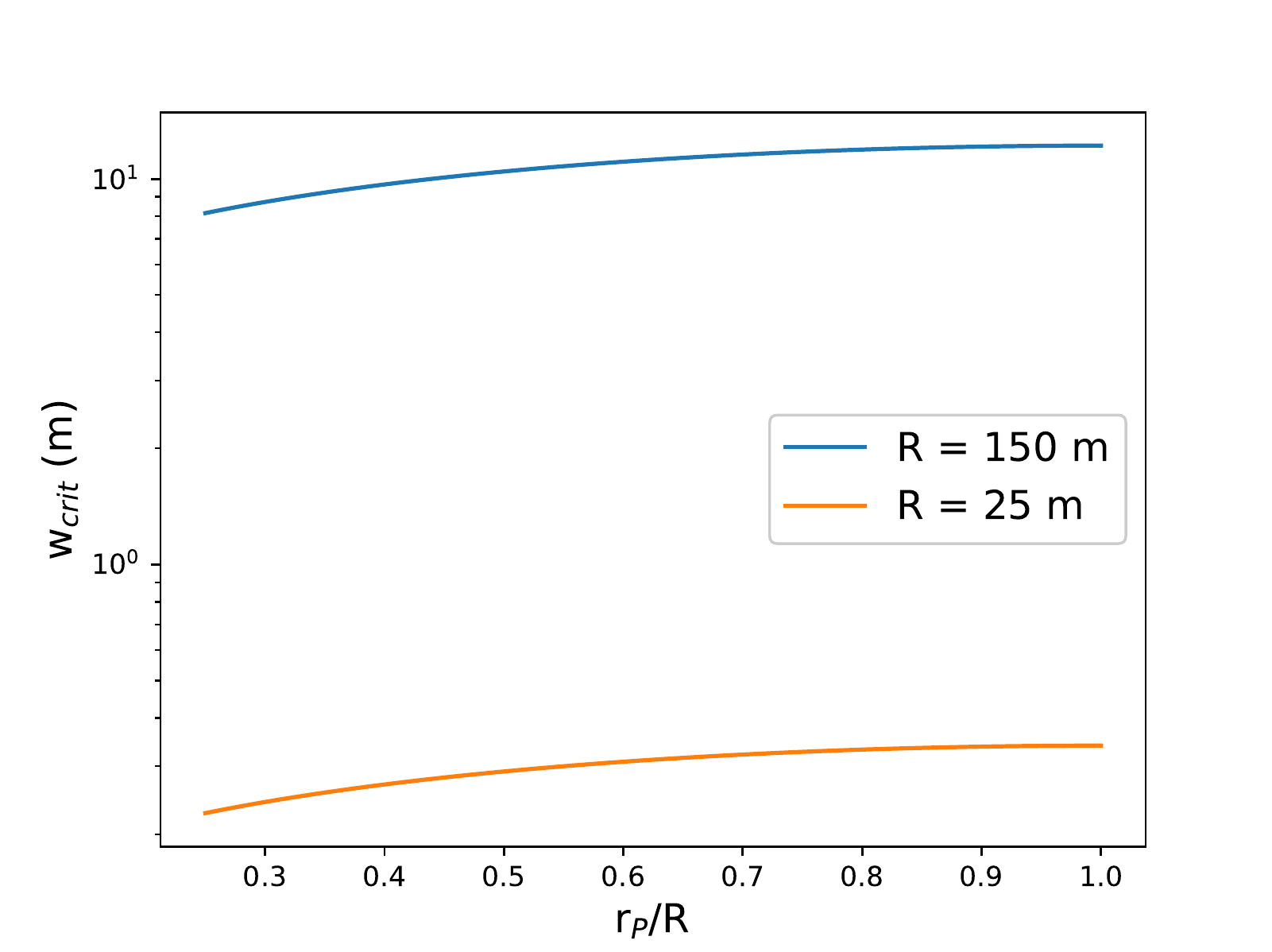}
\caption{Maximum upwards displacement in the centre of an ice sheet just before fracturing, for different sizes $R$, and gas buildup areas $r_{P}$.}
\label{results_w}
\end{figure}

Such a regolith--ice gap may occur due to levitation of the ice slab, but also important is the fact that the slab will bend upwards before fracturing, providing an additional gap for gas to flow through. \citet{Portyankina2010} give an expression for the maximum displacement at the centre of the ice sheet (see Fig.~\ref{fig_model2}), which we can evaluate just before rupture by using the pressure $P_{crit}$ calculated above:
\begin{linenomath*}
\begin{equation}
w_{crit} = \frac{3(1-\nu)r_{P}^{2}P_{crit}}{16Ez^{3}}\left[4(3+\nu)R^{2} - (7+3\nu)r_{P}^{2} - 4 (1+\nu)r_{P}^{2}\ln\frac{R}{r_{P}}\right],
\label{w}
\end{equation}
\end{linenomath*}
with the CO$_{2}$ ice Young's modulus of $E=11.5$ GPa \citep{Yamashita}. The resulting displacement is shown in Figure \ref{results_w}, for the same spider sizes and $r_{P}/R$ ratios as before. The ice in the centre of 25 m and 150 m radius spiders is bent upwards by $\sim20$ cm and $\sim10$ m, respectively, before fracturing occurs. These values greatly exceed the required gap sizes for the gas flow shown in Fig.~\ref{results_freeflow}. Therefore, even though the upwards displacement is smaller towards the edge of the feature, it seems likely that upwards bending of the ice sheet can contribute significantly to the opening of a regolith--ice gap, aiding the flow of gas towards the spider vent.

With viscous flow through regolith pores unable to deliver the full gas production of a reasonable sized spider, and large gas pressure enough to bend or levitate the ice slab, it may be that some spiders undergo both types of flow. For example, low volumes of gas moving through a large cross-sectional area of porous regolith at the outskirts of a feature or around multiple vents, transitioning to free-flow nearer the vent where the converging gas has bent or levitated the overlying ice slab upwards, opening up a gap. Meanwhile, flow through and just above a loose regolith will certainly begin to erode material, digging troughs downwards into it. In \citet{Hao}'s conceptual model, regolith properties are cited as explaining the different appearances and spacing of spider sub-types. Although the above calculations demonstrate that significant gas flow through the pores is unlikely, regolith properties, such as cohesion, could still have an important effect on morphology. So-called `fat spiders' could still be examples of low-cohesion, more easily eroded regoliths, as suggested by \citet{Hao}, for example. Once these troughs and pits are eroded into the regolith, gas will preferentially flow down and further erode them, as described in \citet{Hao}. This will provide additional flow paths for the gas but, since slab ice is thought to conformally cover the terrain in winter, at least some levitation or bending is required to open up a gap again in the spring. The physics of mixed gas flow through the regolith pores, eroded troughs, and a levitated or bent-open ice gap will be complicated, and may involve turbulent as well as viscous flow. Alternatively, local variations in the depth of the impermeable water-ice layer, below the resolution of orbital observations, may allow increased porous gas flow in some localised areas. It seems unlikely, however, that water-ice depths could vary from $\sim$cm to $\sim$metres over a few metres lateral distance in otherwsie identical terrain.

Finally, it should also be noted that some recent work \citep{Chinnery2018} suggests a much lower translucence of CO$_{2}$ ice when compared to previous experiments \citep{Hansen1999}, which would require much thinner ice sheets (on the order of 10 cm) in order to trigger the solid state greenhouse effect. We note that in this case, of a reduced $z$, cryostatic pressure and the critical breaking pressure would be reduced (see Eqn.~\ref{pcrit}), further lowering the pressure gradient and making permeable gas flow even less effective, even for small gas flow rates.

\section{Conclusion}
\label{conclusion}

In the formation of the so--called araneiform features, seen in Mars' southern polar regions, subsurface flow of CO$_{2}$ gas below an ice slab can occur in a gap between ice and regolith, and/or in the substrate itself. The latter possibility has recently received increased attention because spider spatial density scales seem to be related to properties of the underlying regolith. We investigate the effectiveness of subsurface flow of CO$_{2}$ and compare it to gas flow through the gap between ice and regolith. Based on previously estimated flow rates and thermophysical arguments, we suggest that there is insufficient depth of porous regolith to support the full gas flow of all but the very smallest observed spiders through the regolith. On the other hand, free gas flow through a regolith--ice gap is capable of supplying the likely flow rates for gap sizes on the order of a centimetre. This size of gap can be opened in the centre of a spider feature by gas pressure bending the overlying ice slab upwards, or by levitating it entirely as suggested in the original \citet{Kieffer2007} model. Our calculations therefore support at least some of the gas flowing through a gap opened between the regolith and ice. Regolith properties most likely still play a role in the evolution of spider morphology, by regolith cohesion controlling the erosion of the central pit and troughs, for example.

\section*{Acknowledgements}
The authors would like to thank Jingyan Hao and an anonymous reviewer for their constructive and helpful comments which helped improve this manuscript substantially. NA and AH acknowledge support from the UK Space Agency, grant no. ST/R001375/2. EK was supported by STFC grant no. ST/S001271/1.

\bibliography{agujournaltemplate}

\end{document}